\documentclass[a4paper,fleqn,usenatbib]{mnras}

\usepackage[T1]{fontenc}
\usepackage{ae,aecompl}

\usepackage{graphicx}	% Including figure files
\usepackage{amsmath}	% Advanced maths commands
\usepackage{amssymb}	% Extra maths symbols

\newcommand{\eg}{{\em e.g.}}

\newcommand{\sw}{{\it Swift~}}

\newcommand{\xmm}{{\it XMM-Newton~}}
\newcommand{\wse}{{\it WISE~}}

\title[Swift observations of unidentified 3CR sources]{Swift
  observations of unidentified radio sources \\ in the revised Third
  Cambridge Catalogue}

\author[A. Maselli et al.]{
A.~Maselli,$^{1}$
F.~Massaro,$^{2,3,4}$
G.~Cusumano,$^{1}$
V.~La~Parola,$^{1}$
D.~E.~Harris$^{\dagger}$,$^{5}$
\newauthor
A.~Paggi,$^{5}$
E.~Liuzzo,$^{6}$
G.~R.~Tremblay,$^{7}$
S.~A.~Baum,$^{8,9}$
and C.~P.~O'Dea$^{8,10}$
\\
$^{1}$INAF-IASF Palermo, via U. La Malfa 153, I-90146 Palermo, Italy\\
$^{2}$Dipartimento di Fisica, Universit\`a degli Studi di Torino, via Pietro Giuria 1, I-10125 Torino, Italy\\
$^{3}$Istituto Nazionale di Fisica Nucleare, Sezione di Torino, via Pietro Giuria 1, I-10125 Torino, Italy\\
$^{4}$INAF-Osservatorio Astronomico di Roma, Via di Frascati 33, I-00040 Monte Porzio Catone, RM, Italy\\
$^{5}$Harvard-Smithsonian Astrophysical Observatory, 60 Garden Street, Cambridge, MA 02138, USA\\
$^{6}$INAF-Istituto di Radioastronomia, via Gobetti 101, 40129, Bologna, Italy\\
$^{7}$Yale Center for Astronomy and Astrophysics, Physics Department, Yale University, P.O. Box 208120, New Haven, CT 06520-8120, USA\\
$^{8}$University of Manitoba, Dept. of Physics and Astronomy, Winnipeg, MB R3T 2N2, Canada\\
$^{9}$Carlson Center for Imaging Science 76-3144, 84 Lomb Memorial Drive, Rochester, NY 14623, USA\\
$^{10}$School of Physics and Astronomy, Rochester Institute of Technology, 84 Lomb Memorial Drive, Rochester, NY 14623, USA
}

\date{Accepted 2016 May 18. Received 2016 May 18; in original form 2016 March 18}

\pubyear{2016}

\begin{document}
\label{firstpage}
\pagerange{\pageref{firstpage}--\pageref{lastpage}}
\maketitle

\begin{abstract}
We have investigated a group of unassociated radio sources included in
the 3CR catalogue to increase the multi-frequency information on them
and possibly obtain an identification. We have carried out an
observational campaign with the \sw satellite to observe with the UVOT
and the XRT telescopes the field of view of 21 bright NVSS sources
within the positional uncertainty region of the 3CR
sources. Furthermore, we have searched in the recent AllWISE Source
Catalogue for infrared sources matching the position of these NVSS
sources.  We have detected significant emission in the soft X-ray band
for nine of the investigated NVSS sources. To all of them, and in four
cases with no soft X-ray association, we have associated a \wse
infrared counterpart. Eight of these infrared candidates have not been
proposed earlier in the literature. In the five remaining cases our
candidate matches one among a few optical candidates suggested for the
same 3CR source in previous studies. No source has been detected in
the UVOT filters at the position of the NVSS objects, confirming the
scenario that all of them are heavily obscured. With this in mind, a
spectroscopic campaign, preferably in the infrared band, will be
necessary to establish the nature of the sources that we have finally
identified.
\end{abstract}


%%%%%%%%%%%%%%%%%%%%%%%%%%%%%%%%%%%%%%%%%%%%%%%%%%

%%%%%%%%%%%%%%%%% BODY OF PAPER %%%%%%%%%%%%%%%%%%

\section{Introduction}

\let\thefootnote\relax\footnote{$\dagger$~Dan Harris passed away on
  December~6th, 2015. His career spanned much of the history of radio
  and X-ray astronomy. His passion, insight, and contributions will
  always be remembered.}

The extragalactic subset of the revised Third Cambridge Catalogue
(3CR) of radio sources \citep[see,
  \eg,][]{1962MmRAS..68..163B,1985PASP...97..932S} has a long history
as one of the fundamental samples used to understand the nature and
evolution of powerful radio galaxies and quasars, as well as their
relationship to their host galaxies and environments on parsec through
megaparsec scales.
Extensive imaging and spectroscopic observations have long been
available from the radio through the infrared (IR) and optical bands,
with data from {\it Spitzer} \citep{2012ApJ...759...86W}, the {\it
  Hubble Space Telescope}
\citep[\eg,][]{2000A&A...355..873C,2009ApJS..183..278T} and
ground-based telescopes \citep[see, \eg, the description of
  observations performed with the Telescopio Nazionale Galileo
  reported in][]{2009A&A...495.1033B}.

%----------------------------------------------------------------------------------------------------------------------
\begin{table*}
\centering
\scriptsize
\caption{The list of unidentified 3CR sources, with the corresponding NVSS source, when present. 1) The 3C designation; 2-3) Right Ascension (J2000) with its rms uncertainty; 4-5) Declination (J2000) with its rms uncertainty; 6-7) Galactic Longitude and Latitude (J2000); 8) flux density at 178~MHz, corrected following Laing et~al.~(1983); 9-10) corresponding NVSS source, with its flux density at 1.4~GHz; 11) radio spectral index $\alpha$ computed in the range 178~MHz-1.4~GHz.}
\label{tab:3cr}
\begin{tabular}{lcccccccccc}
\hline
~\\                                                                                                                     
3C    &     R.A.     & Error &      Dec.     &  Error   &   l    &     b    & $S_{178}$ &      NVSS        & $S_{1.4}$ & $\alpha$ \\
      &  (hh mm ss)  &  (s)  &  (dd mm ss)   & (arcmin) & (deg)  &   (deg)  &   (Jy)   &                  &   (Jy)  &          \\      
~\\                                                                                                                    
\hline                                                                                                                 
~\\                                                                                                                    
11.1  & 00 29 56.43  & 18.0  & $+$63 40 34.2 &   45.0   & 120.55 &  $+$0.90 &   13.5   & J002945$+$635841 &   2.99  &    0.73  \\ %395   
%~\\                                                                                                                
14.1  & 00 36 27.13  & 18.0  & $+$59 46 30.3 &   45.0   & 121.04 &  $-$3.04 &   17.5   &         -        &    -    &     -    \\ %396   
%~\\                                                                                                                
21.1  & 00 45 35.25  & 18.0  & $+$68 04 23.5 &   45.0   & 122.38 &  $+$5.21 &    9.8   &         -        &    -    &     -    \\ %397   
%~\\                                                                                                                
33.2  & 01 10 19.72  & 18.0  & $+$69 21 57.4 &   60.0   & 124.61 &  $+$6.56 &    6.0   &         -        &    -    &     -    \\ %398   
%~\\                                                                                                                
86    & 03 27 20.10  &  2.0  & $+$55 18 49.7 &    1.0   & 143.91 &  $-$1.08 &   31.6   & J032719$+$552029 &   6.94  &    0.73  \\ %399   
%~\\                                                                                                                
91    & 03 37 42.67  &  2.5  & $+$50 45 45.1 &    3.0   & 147.81 &  $-$3.90 &   15.4   & J033743$+$504552 &   3.34  &    0.74  \\ %400   
%~\\                                                                                                                      
125   & 04 46 16.16  &  5.0  & $+$39 42 23.9 &    7.0   & 164.14 &  $-$3.69 &   15.4   & J044617$+$394503 &   2.02  &    0.98  \\ %401   
%~\\                                                                                                                     
131   & 04 53 22.56  &  3.0  & $+$31 27 47.9 &    3.0   & 171.46 &  $-$7.82 &   15.9   & J045323$+$312924 &   2.87  &    0.83  \\ %402   
%~\\                                                                                                                     
134   & 05 04 42.19  &  1.0  & $+$38 06 12.7 &    1.0   & 167.64 &  $-$1.90 &   81.1   & J050443$+$380539 &   2.14  &    1.05  \\ %403   
%~\\                                                                                                                       
137   & 05 19 32.65  &  3.0  & $+$50 55 40.3 &    3.0   & 158.78 &  $+$7.76 &   13.6   & J051932$+$505432 &   2.07  &    0.91  \\ %404   
%~\\                                                                                                                       
139.2 & 05 24 28.20  &  3.0  & $+$28 13 41.5 &    3.0   & 178.06 &  $-$4.29 &   13.0   & J052427$+$281255 &   0.29  &    0.94  \\ %405   
%~\\                                                                                                                       
141   & 05 26 42.60  &  1.5  & $+$32 49 32.1 &    2.5   & 174.54 &  $-$1.32 &   16.2   & J052642$+$324958 &   2.17  &    0.97  \\ %406   
%~\\                                                                                                                             
152   & 06 04 29.42  &  3.0  & $+$20 21 10.9 &    3.0   & 189.57 &  $-$0.64 &   13.5   & J060428$+$202122 &   1.86  &    0.96  \\ %407   
%~\\                                                                                                                             
158   & 06 21 40.95  &  5.0  & $+$14 29 31.5 &    8.0   & 196.68 &  $+$0.15 &   19.7   & J062141$+$143211 &   2.24  &    1.05  \\ %408   
%~\\                                                                                                                             
250   & 11 08 52.12  &  3.0  & $+$25 00 54.2 &    3.0   & 212.37 & $+$66.91 &    9.6   & J110851$+$250052 &   1.09  &    1.05  \\ %409
%~\\                                                                                                                                       
389   & 18 46 18.63  &  7.0  & $-$03 19 44.5 &   12.0   &  29.38 &  $-$0.38 &   22.9   &         -        &    -    &     -    \\ %410   
%~\\                                                                                                                                  
390   & 18 45 34.41  &  3.0  & $+$09 52 12.9 &    4.0   &  41.08 &  $+$5.77 &   22.9   & J184537$+$095344 &   4.51  &    0.79  \\ %411   
%~\\                                                                                                                                  
394   & 18 59 20.89  &  6.0  & $+$13 00 11.8 &    7.0   &  45.42 &  $+$4.17 &   16.5   & J185923$+$125912 &   2.88  &    0.85  \\ %412   
 %~\\                                                                                                                                 
399.1 & 19 15 56.83  &  3.0  & $+$30 20 02.1 &    3.0   &  62.73 &  $+$8.53 &   14.7   & J191556$+$301952 &   2.97  &    0.78  \\ %413   
%~\\                                                                                                                                  
409   & 20 14 27.74  &  1.0  & $+$23 34 58.4 &    2.0   &  63.40 &  $-$6.12 &   83.5   & J201427$+$233452 &  13.68  &    0.88  \\ %414   
%~\\                                                                                                                                  
415.2 & 20 32 50.51  &  3.0  & $+$53 45 46.1 &    3.0   &  90.27 &  $+$8.19 &    9.6   & J203246$+$534553 &   1.01  &    1.09  \\ %415   
%~\\                                                                                                                                  
428   & 21 08 25.59  &  3.0  & $+$49 34 05.7 &    3.0   &  90.50 &  $+$1.28 &   18.1   & J210822$+$493637 &   2.41  &    0.98  \\ %416   
%~\\                                                                                                                                  
431   & 21 18 55.56  &  3.0  & $+$49 34 18.2 &    3.0   &  91.68 &  $+$0.05 &   26.4   & J211852$+$493658 &   3.39  &    1.00  \\ %417   
%~\\                                                                                                                                  
454.2 & 22 52 15.62  & 18.0  & $+$65 03 57.3 &   45.0   & 110.79 &  $+$5.04 &    9.6   & J225205$+$644010 &   2.29  &    0.69  \\ %418   
%~\\                                                                                                                                  
468.1 & 23 50 54.76  & 18.0  & $+$64 40 19.0 &   45.0   & 116.51 &  $+$2.56 &   32.7   & J235054$+$644018 &   4.95  &    0.92  \\ %419   
~\\                                                                                                                    
\hline  %% rule at top
%~\\
\end{tabular}
\end{table*}
%----------------------------------------------------------------------------------------------------------------------

Since a large fraction of 3CR sources were already present in both the
{\it Chandra} \cite[see, \eg,][for a recent
  review]{2015ApJS..220....5M} and \xmm archives of pointed
observations \cite[\eg][and references therein]{2006ApJ...642...96E},
in 2008 a {\it Chandra} snapshot survey started to complete the X-ray
coverage of the entire 3CR extragalactic catalogue
(\citealt{2010ApJ...714..589M,2012ApJS..203...31M,2013ApJS..206....7M}).
This Chandra survey has enabled investigations of peculiar sources
(see, \eg, \citealp{2009ApJ...696..980M} for 3C~17 and
\citealp{2012MNRAS.419.2338O} for 3C~105), samples of radio-loud
objects \citep{2013ApJ...773...15W,2013ApJ...770..136I}, and was the
genesis of, e.g., follow-up X-ray observations for 3C~89 (Dasadia et
al. 2015), 3C~171 \citep{2010MNRAS.401.2697H}, and 3C~305
\citep{2009ApJ...692L.123M,2012MNRAS.424.1774H}.
The total number of 3CR extragalactic sources now present in the
Chandra archive is 248 out of the 298 included in the update of the
3CR catalogue performed by \cite{1985PASP...97..932S}.
An additional 16 sources (out of 50) that remain unobserved by {\it
  Chandra} have recently been approved for observation in Cycle 17
(see Massaro et~al.~2016).
Those observations began as of December 2015.

Amid our investigation of recent {\it Chandra} observations, we
realised that 25 out of the 298 3CR radio sources are not only
unobserved in X rays, but are in fact completely {\it unidentified},
lacking an assigned optical or infrared counterpart.
In the latest revised release of 3CR extragalactic catalogue
\citep{1985PASP...97..932S}, each of these 25 unidentified sources
(excluding 3C~86 and 3C~415.2) are marked as {\it obscured} active
galaxies.
This classification has remained unchanged for the past three decades,
save for a few tentative associations requiring follow-up observations
for confirmation \cite[see,
  \eg,][]{1987MNRAS.224..847P,1998AJ....115.1348M}.
It therefore became necessary, in nearing completion of the 3CR {\it
  Chandra} snapshot survey, to enact an ancillary optical-to-X-ray
campaign with the \sw observatory in order to better characterise the
properties of these unidentified sources.
Our \sw campaign was augmented by a search for infrared counterparts
in the latest AllWISE Source
Catalogue \footnote{\underline{http://wise2.ipac.caltech.edu/docs/release/allsky/}}
from the Wide-field Infrared Survey Explorer (WISE,
\citealt{2010AJ....140.1868W}) mission.

%---------------------------------------------------------------------------------------------------------------------------------------------------------------------------------
\begin{table*}
\centering
\scriptsize
\caption{The list of \sw-XRT detected sources matching one of the NVSS sources listed in Table~\ref{tab:3cr}. 1) The 3C designation; 2-3) Equatorial coordinates of the X-ray source; 4) error radius; 5) XRT exposure time; 6) XRT count rate with its error; 7) significance of the X-ray detection; 8) corresponding NVSS source; 9) angular separation between the X-ray and the radio source.}
\label{tab:X}
\begin{tabular}{lcccccccc}
\hline
~\\                                                                                                                        
 3C   & R.A.(J2000) & Dec. (J2000)  &   Error  & Exposure &            Count Rate          &     S/N    &      NVSS        & Angular Separation \\
      &             &               & (arcsec) &    (s)   &              (ct/s)            & ($\sigma$) &                  &     (arcsec)      \\     
~\\                                                                                                                                     
\hline                                                                                                                                  
~\\                                                                                                                                     
 86   &  03 27 19.5 & $+$55 20 26.0 &    4.5   &   5170   & (1.43+/-0.19) $\cdot$ 10$^{-2}$ &     7.6    & J032719$+$552029 &        3.7        \\
%~\\                                                                                                                                    
 91   &  03 37 43.0 & $+$50 45 46.2 &    4.0   &   4809   & (4.39+/-0.39) $\cdot$ 10$^{-2}$ &    11.3    & J033743$+$504552 &        7.4        \\
%~\\                                                                                                                                    
131   &  04 53 23.2 & $+$31 29 33.4 &    6.3   &   5255   & (1.96+/-0.74) $\cdot$ 10$^{-3}$ &     2.6    & J045323$+$312924 &        9.4        \\ 
%~\\                                                                                                                            
137   &  05 19 32.6 & $+$50 54 31.4 &    4.7   &   5092   & (9.98+/-1.60) $\cdot$ 10$^{-3}$ &     6.1    & J051932$+$505432 &        1.9        \\
%~\\                                                                                                                            
158   &  06 21 41.2 & $+$14 32 11.5 &    6.4   &   4974   & (2.04+/-0.78) $\cdot$ 10$^{-3}$ &     2.6    & J062141$+$143211 &        1.6        \\
%~\\                                                                                                                            
390   &  18 45 37.6 & $+$09 53 48.7 &    4.4   &   4348   & (2.60+/-0.30) $\cdot$ 10$^{-2}$ &     8.6    & J184537$+$095344 &        4.4        \\
%~\\                                                                                                                            
409   &  20 14 27.5 & $+$23 34 54.5 &    4.0   &   5366   & (3.27+/-0.28) $\cdot$ 10$^{-2}$ &    11.6    & J201427$+$233452 &        1.9        \\
%~\\                                                                                                                            
428   &  21 08 22.1 & $+$49 36 42.1 &    4.6   &   7913   & (7.79+/-1.10) $\cdot$ 10$^{-3}$ &     6.9    & J210822$+$493637 &        5.6        \\
%~\\                                                                                                                            
454.2 &  22 52 05.2 & $+$64 40 13.1 &    4.6   &   5571   & (7.76+/-1.40) $\cdot$ 10$^{-3}$ &     5.6    & J225205$+$644010 &        4.6        \\ 
~\\                                                                                                                                             
\hline  %% rule at top
~\\                                                                                                                                             
\end{tabular}
%\tablefoot{}
\end{table*}
%----------------------------------------------------------------------------------------------------------------------------------------------------------------------

Here we present the results of this new observational effort.
Our \sw observing strategy is described in \S~\ref{sec:nvss}, the
reduction and analysis of \sw X-ray data is discussed in
\S~\ref{sec:obs}, and the search for infrared and optical counterparts
is described in \S~\ref{sec:lower}.
Our results, including detections of both infrared and soft X-ray
counterparts, are given in \S~\ref{sec:details} and summarised in
\S~\ref{sec:summary}.
Throughout this paper we use CGS units, unless stated otherwise. 
The spectral index $\alpha$ is defined in terms of the flux density
S$_{\nu}$, where S$_{\nu}\propto\nu^{-\alpha}$ and $\nu$ is the
frequency.

\section{Observing Strategy}
\label{sec:nvss}

The coordinates of the 3CR sources were first provided by
\cite{1962MmRAS..68..163B} and later modified up to the most recent
update carried out by \cite{1985PASP...97..932S}.
In several cases, including a few sources of interest, the positional
uncertainty reaches values up to 60~arcmin in Declination.
Given the high flux density threshold used in selecting sources for
the 3CR catalogue, we have assumed that a bright source in the more
recent NRAO VLA Sky Survey \cite[NVSS,][]{1998AJ....115.1693C} at
1.4~GHz would be associated with all unidentified 3CR sources.
The positional uncertainty of NVSS objects with flux density values
higher than 100~mJy is always lower than one arcsecond.
The 3CR catalogue included sources with flux density values $S_{178}$
higher than 9~Jy: considering the radio spectral index distribution of
unidentified 3CR sources reported by \cite{1968AJ.....73..953P}, we
have established a lower limit of $S_{1.4}^{\star} = 1$~Jy for the
expected flux density at 1.4~GHz.
Therefore, to establish the most suited coordinates to be used in our
\sw campaign, we have searched for NVSS sources with $S_{1.4} >
S_{1.4}^{\star}$.

In most cases the result of our search was a single NVSS source with a
compact radio morphology.
In two cases (for 3C~134 and 3C~139.2) we found a group of three
catalogued objects that in actuality correspond to the same radio
source - both of these are radio galaxies with Fanaroff-Riley class II
\cite[FR~II,][]{1974MNRAS.167P..31F} morphology
\citep{1984MNRAS.210..929L}, that NVSS has spatially resolved into a
radio core and two jet hotspots.
For both of these FR~II sources (see, \eg, Fig.\ref{fig:405}) only one
of the NVSS sources is internal to the 3CR positional uncertainty
region.
We note that for 3C~139.2 the flux density of the NVSS object
(presumably the radio core) within the 3CR positional uncertainty
region is lower than $S_{1.4}^{\star}$, but we have nevertheless
included it in our observational campaign considering the contribution
of the two hotspots (0.7~Jy and 0.9~Jy, respectively) to the total
flux.
We have also decided to include three NVSS objects with an angular
separation of a few arcseconds from the closest side of the
corresponding 3CR positional uncertainty region: 1.8 and 3.4~arcsec
for 3C~390 and 3C~428, respectively, and $\sim$39~arcsec for 3C~86.
In these cases, as a consequence of this small angular separation, the
NVSS radio contours (up to 10~mJy~beam$^{-1}$) largely overlap the
corresponding 3CR positional uncertainty region.
On the other hand, we have excluded 3C~14.1, 3C~21.1, 3C~33.2, and
3C~389 from our initial dataset as we have found no bright NVSS source
in or near (within several arcminutes) the corresponding 3CR
positional uncertainty region.

The list of these 21 previously unassociated 3CR sources is given in
Table~\ref{tab:3cr}.
For each, we report the Equatorial coordinates with their statistical
uncertainty, the Galactic coordinates, their flux density at 178~MHz,
the corresponding NVSS source with its flux density at 1.4~GHz, and
the radio spectral index $\alpha$ computed between 178~MHz and
1.4~GHz.
The flux density values at 178~MHz include a 9~per~cent correction
factor \citep{1983MNRAS.204..151L} to those originally provided by
\cite{1969ApJ...157....1K} and also reported by
\cite{1985PASP...97..932S}.
For both 3C~134 and 3C~139.2 only the NVSS object internal to the 3CR
positional uncertainty region has been reported in
Table~\ref{tab:3cr}; however, in the computation of the radio spectral
index $\alpha$ we have included the contribution of the additional
NVSS objects that correspond to the same radio galaxy, under the
assumption that the older survey at 178~MHz was unable to spatially
resolve them.

\section{X-ray Data Reduction and Analysis Procedures}
\label{sec:obs}

Each of these 21 NVSS sources was covered by our \sw observational
campaign, carried out between November~2014 and March~2015, with a
total exposure time greater than 4~ks for each source.
The X-ray data reduction and procedures adopted in the present
analysis were extensively described in
\cite{2008A&A...479...35M,2008A&A...489.1047M,2013ApJS..209....9P},
and references therein; here we report only the basic details (see
also \citealp{2004SPIE.5165..217H,2005SSRv..120..165B} for further
details).

The XRT data have been processed with the XRTDAS software package
(v.3.0.0) developed at the ASI Science Data Center (ASDC) and
distributed within the HEASoft package (v.6.16) by the NASA High
Energy Astrophysics Archive Research Center (HEASARC).
All the XRT observations were carried out in the most sensitive photon
counting (PC) readout mode.
Event files have been calibrated and cleaned applying standard
filtering criteria with the {\sc xrtpipeline} task and using the
latest calibration files available in the Swift CALDB distributed by
HEASARC.
Events in the energy range 0.3--10 keV with grades 0--12 have been
used in the analysis.
Exposure maps have been also created with {\sc xrtpipeline}.
The detection of X-ray sources in the XRT images has been carried out
using the detection algorithm {\sc detect} within {\sc ximage}.
In agreement with \cite{2011A&A...528A.122P}, we have set the {\sc
  detect} signal-to-noise ratio acceptance threshold to 2.5~$\sigma$.
Finally, the positional uncertainty (90 per cent confidence level) of
each detected source has been computed using the {\sc xrtcentroid}
task.
When needed, we have computed a 3$\sigma$ upper limit of the count
rate at the desired coordinates using the {\sc uplimit} command within
{\sc ximage}.
Source count rate photometry was conducted using square boxes with
half-size of 7 pixels, while background intensity was set to a
constant equal to the average value computed over the whole image.

The list of XRT sources that we have detected following the above
described procedure and matching one NVSS source is reported in
Table~\ref{tab:X}, for a total of nine X-ray sources.
For each, we report the corresponding 3CR source (column~1), its Right
Ascension (column 2) and Declination (column 3), the error radius
(column~4), the XRT exposure time (column~5), the count rate with its
error (column~6), the significance of its detection (column~7), the
corresponding NVSS source (column~8) and the angular separation from
its position (column~9).
The name adopted for these sources in this paper starts with the
prefix XRT and then encodes the J2000 sky position following the
standard IAU convention (\eg, XRT~JHHMMSS.S$+$DDMMSS).
In three cases (3C~91, 3C~131 and 3C~428) the NVSS source has been
found at an angular separation higher than the corresponding XRT error
radius (see the exact values reported in Table~\ref{tab:X}).
In principle, a marginal disagreement between the positions of the
catalogued radio source, which is really the radio centroid weighted
in position by any intrinsic asymmetry in the radio structure, and of
the detected X-ray source (which should mark the AGN position more
precisely) is not unexpected.
Considering this, and the fact that the XRT error radius is given at a
90 per cent confidence level, in the first instance we have not
rejected the match and we have later verified the possible agreement
of the X-ray object with the position of a \wse object (see
Section~\ref{sec:lower}) rather than the NVSS one.
Further detected X-ray sources were found at much higher angular
separations from the catalogued NVSS objects, so that any relation
between them could be safely excluded.
Regarding the twelve NVSS sources for which no significant (higher
than 2.5$\sigma$) X-ray detection has been found, a 3~$\sigma$ upper
limit has been computed at the position corresponding to the NVSS
coordinates; the list of these is reported in Table~\ref{tab:noX}.

%---------------------------------------------------------------------------------------------------------------------------------------------------------------------------------
\begin{table}
\centering
\scriptsize
\caption{The list of the NVSS sources with no matching X-ray detection and the corresponding \sw-XRT 3$\sigma$ upper limit.}
\label{tab:noX}
\begin{tabular}{lccc}
\hline
~\\                                                                                                                        
  3C   &       NVSS       & Exposure & 3~$\sigma$ upper limit \\
       &                  &    (s)   &        (ct/s)          \\     
~\\                                                              
\hline                                                          
~\\                                                              
  11.1 & J002945$+$635841 &   5263   & 1.69 $\cdot$ 10$^{-3}$ \\
%~\\                                                  
 125   & J044617$+$394503 &   4132   & 4.41 $\cdot$ 10$^{-3}$ \\
%~\\                                             
 134   & J050443$+$380539 &   4944   & 3.46 $\cdot$ 10$^{-3}$ \\ 
%~\\                                             
 139.2 & J052427$+$281255 &   4817   & 3.21 $\cdot$ 10$^{-3}$ \\ 
%~\\                                                                                                                      
 141   & J052642$+$324958 &   4651   & 2.24 $\cdot$ 10$^{-3}$ \\
%~\\                                                                                                                      
 152   & J060428$+$202122 &   4894   & 1.87 $\cdot$ 10$^{-3}$ \\ 
%~\\                                             
 250   & J110851$+$250052 &   4280   & 4.80 $\cdot$ 10$^{-3}$ \\
%~\\                                             
 394   & J185923$+$125912 &   4032   & 5.32 $\cdot$ 10$^{-3}$ \\   
%~\\                                       
 399.1 & J191556$+$301952 &  12339   & 2.20 $\cdot$ 10$^{-3}$ \\
%~\\                                             
 415.2 & J203246$+$534553 &   5358   & 3.59 $\cdot$ 10$^{-3}$ \\
%~\\                                             
 431   & J211852$+$493658 &   7462   & 2.77 $\cdot$ 10$^{-3}$ \\
%~\\                                             
468.1  & J235054$+$644018 &   4997   & 3.10 $\cdot$ 10$^{-3}$ \\
~\\                                                                                                                                             
\hline  %% rule at top
\end{tabular}
%\tablefoot{}
\end{table}
%----------------------------------------------------------------------------------------------------------------------------------------------------------------------

%---------------------------------------------------------------------------------------------------------------------------------------------------------------------------------
%\setcounter{table}{1}                                                         
\begin{table*}
\centering
\scriptsize
\caption{The \wse counterparts associated to some of the NVSS sources reported in Table~\ref{tab:X}.}
\label{tab:lower}
\begin{tabular}{lcccccccc}
\hline
~\\                                                                                                                        
3C   &     NVSS         &           \wse              &      WISE/NVSS     &   WISE/XRT   &      $w1$        &       $w2$       &       $w3$       &      $w4$        \\
     &                  &                             &      (arcsec)      &   (arcsec)   &      (mag)       &       (mag)      &       (mag)      &      (mag)       \\
~\\                                                                                             
\hline                                                                                         
~\\                                                                                            
86    & J032719$+$552029 &      J032719.29$+$552028.2  &        0.9         &     2.7      & 13.421$\pm$0.027 & 12.500$\pm$0.026 &  9.757$\pm$0.050 & 7.274$\pm$0.128 \\
%~\\                                                                                   
91    & J033743$+$504552 &      J033743.02$+$504547.6  &        6.1         &     1.4      & 11.885$\pm$0.022 & 10.802$\pm$0.021 &  7.936$\pm$0.020 & 5.507$\pm$0.038 \\
%~\\                                                                                         
131   & J045323$+$312924 &      J045323.34$+$312928.4  &        4.0         &     5.4      & 14.981$\pm$0.041 & 14.779$\pm$0.082 & 12.306           & 8.300           \\
%~\\                                                                          
137   & J051932$+$505432 &      J051932.53$+$505431.3  &        1.5         &     0.7      & 13.967$\pm$0.027 & 12.896$\pm$0.028 &  9.968$\pm$0.053 & 7.194$\pm$0.116 \\
%~\\                                                                                              
158   & J062141$+$143211 &      J062141.01$+$143212.8  &        1.5         &     3.6      & 15.133$\pm$0.046 & 13.953$\pm$0.043 & 11.131$\pm$0.189 & 8.639$\pm$0.417 \\
%~\\                                                                                              
390   & J184537$+$095344 &      J184537.60$+$095345.0  &        0.9         &     3.8      & 12.546$\pm$0.043 & 11.575$\pm$0.024 &  9.150$\pm$0.029 & 6.874$\pm$0.088 \\
%~\\                                                                                              
409   & J201427$+$233452 &      J201427.59$+$233452.6  &        0.3         &     2.1      & 13.547$\pm$0.050 & 12.377$\pm$0.027 &  9.005$\pm$0.027 & 6.437$\pm$0.065 \\
%~\\                                                                                              
428   & J210822$+$493637 &      J210822.08$+$493641.6  &        5.3         &     0.5      & 14.559$\pm$0.064 & 13.097$\pm$0.035 & 10.143$\pm$0.056 & 7.601$\pm$0.136 \\
%~\\                                                                                              
454.2 & J225205$+$644010 &      J225205.50$+$644011.9  &        2.3         &     4.6      & 14.652$\pm$0.030 & 14.341$\pm$0.042 & 13.121$\pm$0.467 & 9.513           \\
~\\                                                                                                 
\hline                                                                                              
~\\                                                                                                 
125   & J044617$+$394503 &      J044617.88$+$394504.5  &        1.6         &     ---      & 15.645$\pm$0.052 &	15.228$\pm$0.092 & 10.605$\pm$0.100 & 7.825$\pm$0.189 \\
%~\\                                                                                              
139.2 & J052427$+$281255 &      J052427.51$+$281256.7  &        1.5         &     ---      & 15.782$\pm$0.060 &	14.954$\pm$0.086 &  9.475$\pm$0.046 & 5.987$\pm$0.051 \\
%~\\                                                                                              
152   & J060428$+$202122 &      J060428.62$+$202121.7  &        0.8         &     ---      & 16.334$\pm$0.096 &	16.576$\pm$0.337 & 12.245$\pm$0.450 & 8.229 	      \\
%~\\                                                                                              
468.1 & J235054$+$644018 &      J235054.78$+$644018.1  &        1.3         &     ---      & 15.135$\pm$0.040 & 13.452$\pm$0.029 &  9.946$\pm$0.044 & 7.966$\pm$0.138 \\
~\\                                                                                              
\hline  %% rule at top
\end{tabular}
%\tablefoot{}
\end{table*}
%----------------------------------------------------------------------------------------------------------------------------------------------------------------------

\section{Search for infrared and optical counterparts}
\label{sec:lower}

We have searched for infrared counterparts to the NVSS sources
reported in Table~\ref{tab:3cr} by cross-matching this list with the
AllWISE Source Catalogue.
Following \cite{2013ApJS..206...12D}, we have used the value of
3.3~arcsec as proper matching radius.
In the previously cited cases of 3C~91, 3C~131 and 3C~428 (see
Section~\ref{sec:obs}), an infrared object has been found at an
angular separation higher than 3.3~arcsec from the NVSS object (exact
values are shown in Table~\ref{tab:lower}, described below), which is
nevertheless still within the error circle of the corresponding XRT
source.
Having found positional agreement between the infrared and the soft
X-ray objects, with only a modest difference with respect to the NVSS
coordinates, we accept these objects as counterparts of the same
source.
As a result of our analysis, an infrared counterpart has been found
for all sources reported in Table~\ref{tab:X}, as well as in four
additional cases among the X-ray non-detections listed in
Table~\ref{tab:noX}.
Their list is reported in Table~\ref{tab:lower}, according to the
presence/absence of an X-ray counterpart.
For each infrared source we have reported the corresponding 3CR
(column~1) and NVSS (column~2) sources, the name in the AllWISE
Catalogue (column~3), the angular separation from the radio (column~4)
and the X-ray source (column~5), and the corresponding magnitudes in
the \wse filters (columns~6--9).

We have also carried out a photometric analysis over the images in the
available UVOT filters at the position of the NVSS sources.
Following \cite{2013ApJS..206...17M}, the photometry has been
performed using the {\sc uvotdetect} task and taking into account the
corresponding exposure maps.
An extraction region of 5~arcsec has been adopted for the sources,
independently of the image filter, and a larger circle of 20~arcsec
radius for the background, in a near source-free region of the sky.
Magnitude values, or upper limits, in the Vega System have been
finally obtained using the {\sc uvotsource} task and adopting a
3$\sigma$ level of significance to compute the background limit; both
statistic and systematic errors have been taken into account.
Our photometric results show that an optical-UV counterpart has not
been detected for any object; only upper limits could be established.
This result was not unexpected, considering that almost all of these
sources were marked as {\it obscured} by \cite{1985PASP...97..932S},
and nearly all are at low Galactic latitude ($\mid b \mid
<10^{\circ}$, see Table~\ref{tab:3cr}).

\section{Source Details}
\label{sec:details}

\begin{figure}
\begin{center}
\includegraphics[height=3.9cm]{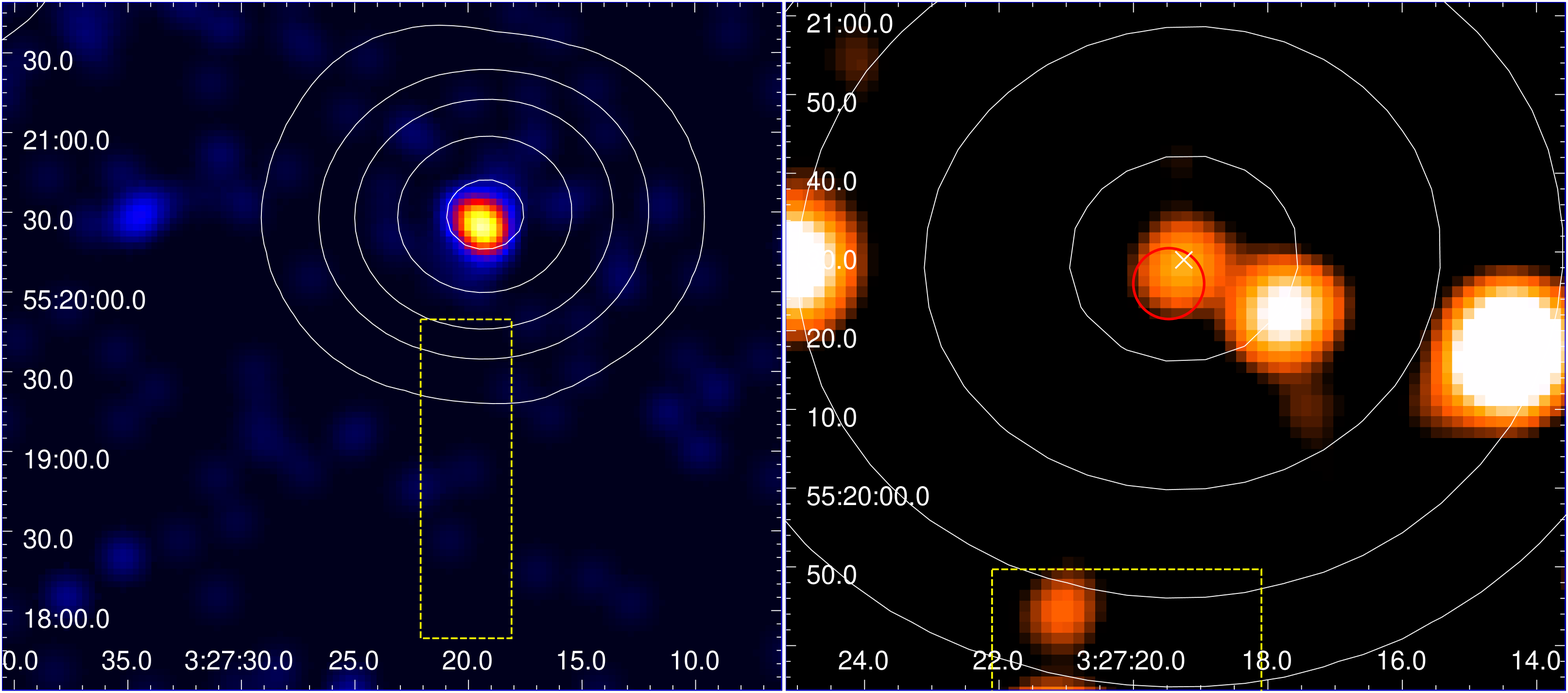}
\end{center}
\caption{The sky map in the direction of {\bf 3C~86} obtained by XRT
  in the 0.3--10 keV energy band (left panel) and by \wse in the $w1$
  filter (right panel). A yellow dashed line marks the positional
  uncertainty region of the 3CR source. White continuous lines shape
  the radio contours obtained from the NVSS map and corresponding to
  0.01, 0.2, 0.7, 2, and 4~Jy~beam$^{-1}$; a white cross marks the
  position of the catalogued NVSS source. A red circle marks the
  position of the detected XRT source with the corresponding error
  radius.}
\label{fig:399}
\end{figure}

\begin{figure}
\begin{center}
\includegraphics[height=3.9cm]{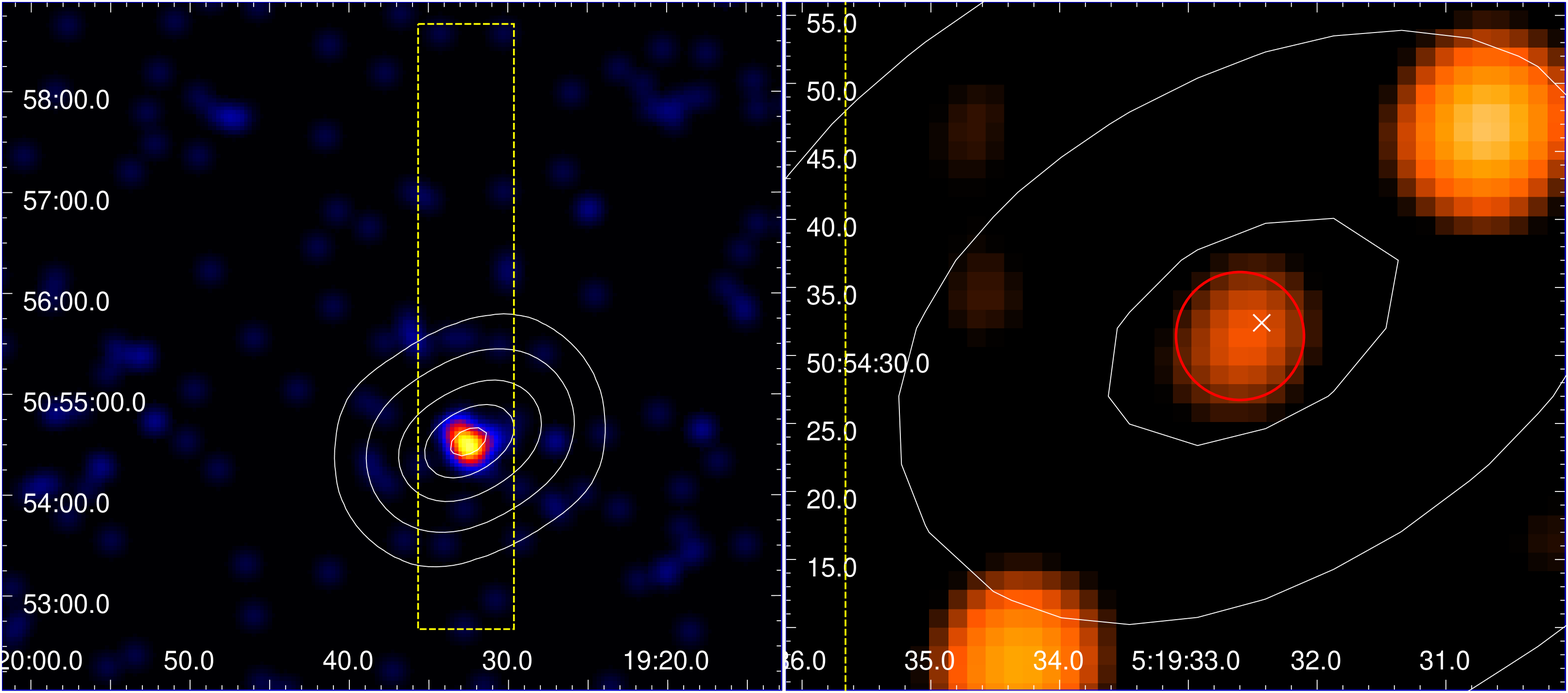}
\end{center}
\caption{Same as Fig.~\ref{fig:399} but for {\bf 3C~137}. White
  contours correspond in this case to 0.01, 0.1, 0.4, 0.8, and
  1~Jy~beam$^{-1}$.}
\label{fig:404}
\end{figure}

\begin{figure}
\begin{center}
\includegraphics[height=3.9cm]{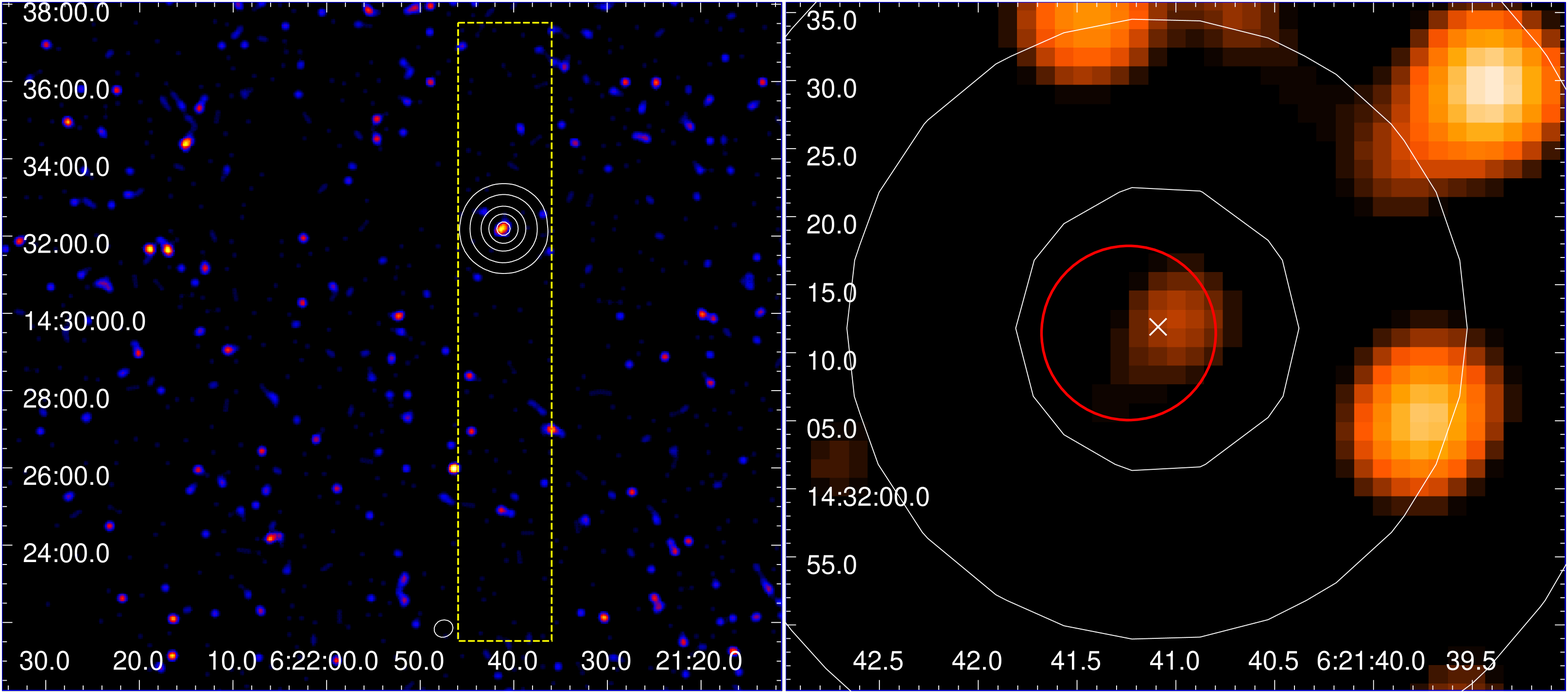}
\end{center}
\caption{Same as Fig.~\ref{fig:399} but for {\bf 3C~158}. White
  contours correspond in this case to 0.01, 0.1, 0.5, 1, and
  1.5~Jy~beam$^{-1}$.}
\label{fig:408}
\end{figure}

\begin{figure}
\begin{center}
\includegraphics[height=3.9cm]{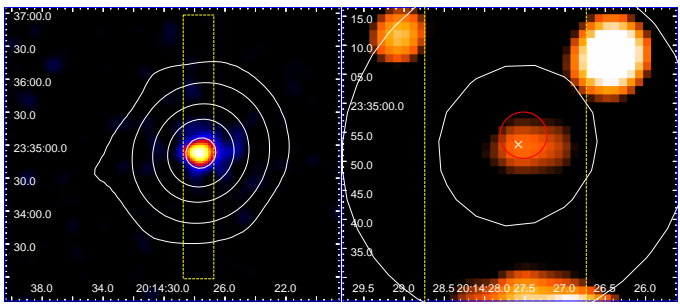}
\end{center}
\caption{Same as Fig.~\ref{fig:399} but for {\bf 3C~409}. White
  contours correspond in this case to 0.01, 0.2, 1.5, 4, and
  8~Jy~beam$^{-1}$.}
\label{fig:414}
\end{figure}

In this Section we describe details of the 13 radio sources listed in
Table~\ref{tab:lower}, distinguishing the presence
(Section~\ref{X_ir}) or the absence (Section~\ref{only_ir}) of an
X-ray counterpart in addition to the infrared one.
For each radio source we show a comparison of the field of view in
both the XRT 0.3--10 keV band and in \wse $w1$ filter
(figures~\ref{fig:399} to~\ref{fig:419}).
The images have been smoothed with a Gaussian function with different
values of FWHM (5 and 1~arcsec for the XRT and \wse images,
respectively).
In each panel a yellow dashed line marks the positional uncertainty
region of the 3CR source.
White crosses mark the position of the catalogued NVSS objects, and
white continuous lines are used to shape the radio contours that have
been obtained from the NVSS maps.
The exact values of the contour levels, starting from
10~mJy~beam$^{-1}$, have been reported for each in the corresponding
captions.
The positions of the X-ray sources and the corresponding error radius
(see column~3 of Table~\ref{tab:X}) have been marked with red circles.
The range of the image in the XRT band (left panel) has been generally
chosen to cover the whole 3CR positional uncertainty region and the
corresponding NVSS source with its contours; in two cases (3C~454.2
and 3C~468.1) this was not convenient due to the large extent in
Declination of the positional uncertainty region.
The image in the \wse $w1$ filter shows a smaller field of view in
order to aid viewing of the possible infrared counterpart.

\subsection{3CR sources with both X-ray and infrared counterparts}
\label{X_ir}

\subsubsection{3C 86}
\label{sub:399}

As shown in the left panel of Fig.~\ref{fig:399} the 1.4~GHz source
corresponding to 3C~86, NVSS~J032719$+$552029 (S$_{1.4}$=6.9~Jy), is
out of the 3CR positional uncertainty region but its radio contours
overlap it.
We have detected X-ray emission (XRT~J032719.5$+$552026) cospatial
with the radio source at $S/N$=7.6 and with a mean count rate of the
order of 10$^{-2}$~ct/s.
Furthermore, at an angular separation of 0.9~arcsec from the NVSS
coordinates, an infrared counterpart \wse~J032719.29$+$552028.2 has
been found in the AllWISE Source Catalogue with clear detections in
all four \wse filters.
This is the same counterpart reported by \cite{1991A&A...244...37H} in
their investigation of three 3CR sources including a spectroscopic
analysis: unfortunately, the spectrum they obtained had very low
signal-to-noise ratio, showing only faint continuum.

\begin{figure}
\begin{center}
\includegraphics[height=3.9cm]{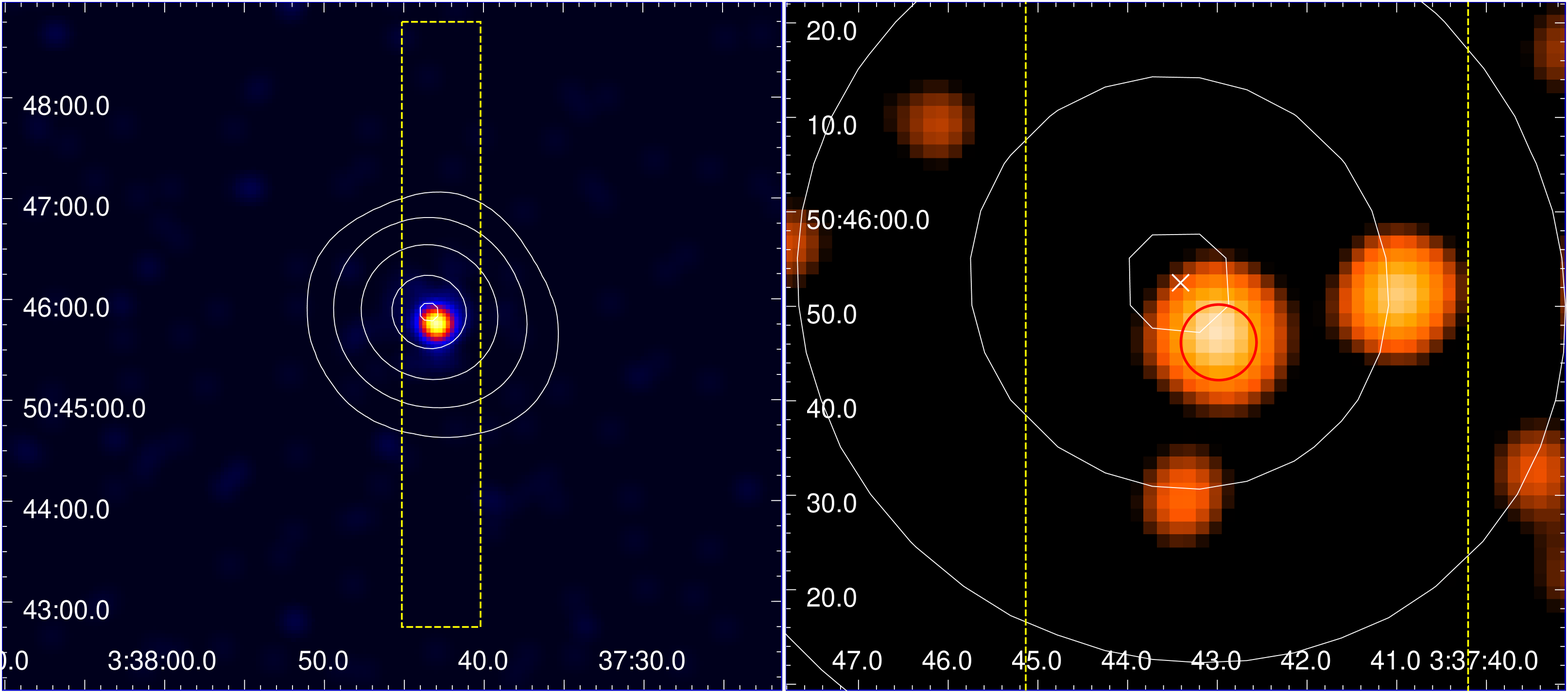}
\end{center}
\caption{Same as Fig.~\ref{fig:399} but for {\bf 3C~91}. White
  contours correspond in this case to 0.01, 0.1, 0.5, 1.5, and
  2.3~Jy~beam$^{-1}$.}
\label{fig:400}
\end{figure}

\subsubsection{3C 91}
\label{sub:400}

The X-ray source that we have detected (XRT~J033743.0$+$504546) with
$S/N$=11.3 and a mean count rate on the order of
4~$\cdot$~10$^{-2}$~ct/s is 7.4~arcsec from the coordinates of
NVSS~J033743$+$504552 and 3.3~arcsec from the center of the 3CR
positional uncertainty region, as shown in the left panel of
Fig.~\ref{fig:400}.
A \wse source, J033743.02$+$504547.6, has been found within the XRT
error circle, at close angular separation (1.4~arcsec) from its
center, and is clearly detected in all \wse filters.
Considering the good positional agreement between the infrared and the
X-ray objects and their low angular separation from the NVSS object
(see the right panel of Fig.~\ref{fig:400}), we have accepted these as
counterparts of the same source at different frequencies.
Three candidates were noted by \cite{1998AJ....115.1348M} in their
analysis of the optical images obtained by the Wide Field Planetary
Camera 2 (WFPC2) on board the {\it Hubble Space Telescope} (HST).
The angular separation between their candidate \#1 (R.A. 03$^{h}$
37$^{m}$ 42.93$^{s}$; Dec. $+$50$^{\circ}$ 45$^{'}$ 48.13$^{''}$),
which they considered as the most probable, and
\wse~J033743.02$+$504547.6 is 1.0~arcsec.
\cite{1998AJ....115.1348M} reported for their candidate \#1 an
observed magnitude of R$_{obs}$=19.36~mag; there is no detection in
the available UVOT filters ($M2$, $W1$, and $W2$) at the corresponding
position.
Finally, we report that \wse~J033743.02$+$504547.6 has been included
by \cite{2014ApJS..215...14D} in their all-sky catalogue of infrared
selected, radio-loud active galaxies due to its peculiar infrared
colours.

\begin{figure}
\begin{center}
\includegraphics[height=3.9cm]{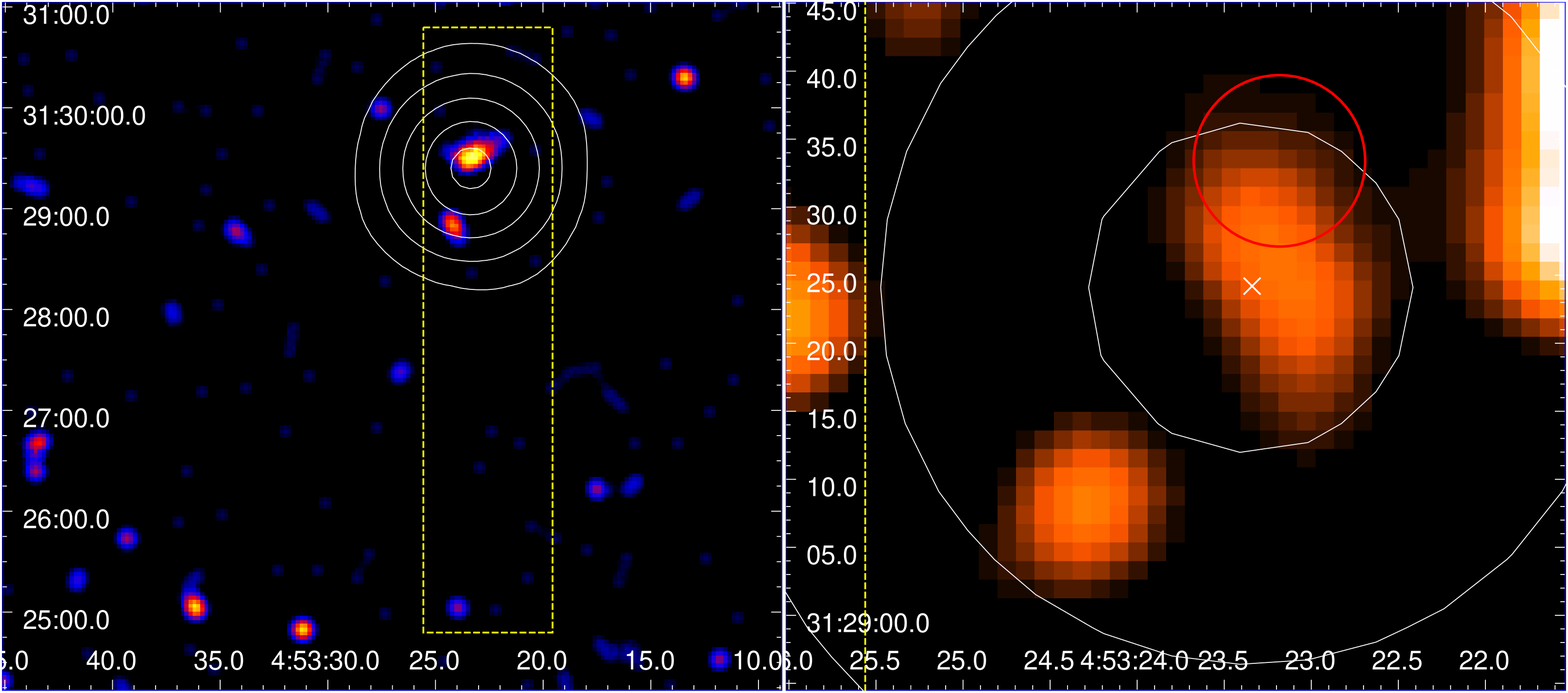}
\end{center}
\caption{Same as Fig.~\ref{fig:399} but for {\bf 3C~131}. White
  contours correspond in this case to 0.01, 0.1, 0.4, 1, and
  1.8~Jy~beam$^{-1}$.}
\label{fig:402}
\end{figure}

\subsubsection{3C 131}
\label{sub:402}

The X-ray source XRT~J045323.2$+$312933 has been detected at
9.4~arcsec from the coordinates of NVSS~J045323$+$312924.
There are two \wse objects at close angular separation
($\sim$4~arcsec) from the NVSS source: in the image shown in
Fig.~\ref{fig:402} (right panel) they are not resolved, and reliable
magnitude values of both targets are only available for the $w1$ and
$w2$ filters in the AllWISE Source Catalogue.
Only one, \wse~J045323.34$+$312928.4, is within the error circle of
the XRT source.
No optical candidate counterpart was supported by
\cite{1987MNRAS.224..847P} and the only cited source was considered to
be unrelated to the radio structure.
This region of the sky was also analysed by
\cite{1998AJ....115.1348M}, who reported a list of four objects
detected by HST.
The angular separation of their candidate \#4 (R.A. 04$^{h}$ 53$^{m}$
23.34$^{s}$; Dec. $+$31$^{\circ}$ 29$^{'}$ 27.10$^{''}$) from
\wse~J045323.34$+$312928.4 is 1.3~arcsec.
Due to their large angular separation from the radio coordinates they
took into account, different from the NVSS ones, all the four
candidates were considered by \cite{1998AJ....115.1348M} to be
unlikely the optical counterpart of the radio source.
However, the angular separation of candidate \#4 from
NVSS~J045323$+$312924 is 2.9~arcsec, lower than the positional
uncertainty for the HST coordinates (3~arcsec) quoted by
\cite{1998AJ....115.1348M}.
Therefore, despite a difference of a few arcseconds (exact values are
reported in Table~\ref{tab:X} and Table~\ref{tab:lower}) between the
positions of objects detected at different frequencies, we suggest
that the most plausible counterpart to the radio source 3C~131
corresponds to \wse~J045323.34$+$312928.4.

\subsubsection{3C 137}
\label{sub:404}

The X-ray source XRT~J051932.6$+$505431 matches the coordinates of the
NVSS source J051932$+$505432 with an angular separation of 1.9~arcsec.
Also a reliable infrared counterpart, \wse~J051932.53$+$505431.3, is
found at 1.5~arcsec from the NVSS coordinates and is well detected in
all the \wse filters.
The angular separation of this infrared source from the "very faint
red object", quoted by \cite{1987MNRAS.224..847P} in their search of
an optical identification and hardly distinguished in their finding
chart, is 3.4~arcsec.

\subsubsection{3C 158}
\label{sub:408}

The X-ray source XRT~J062141.2$+$143212 matches the NVSS source
J062141$+$143211, within the positional uncertainty region of 3C~158,
with an angular separation of 1.6~arcsec consistent with the XRT error
circle.
An infrared counterpart in the AllWISE Source Catalogue,
\wse~J062141.01$+$143212.8, is well detected in all the remaining \wse
filters and is found at 1.5~arcsec from the NVSS source.
This object, with a nice positional agreement with other sources
emitting at different frequencies, is here presented for the first
time as candidate to be investigated with a spectroscopic analysis; no
infrared/optical candidate has been previously reported in the
literature for this radio source.

\begin{figure}
\begin{center}
\includegraphics[height=3.9cm]{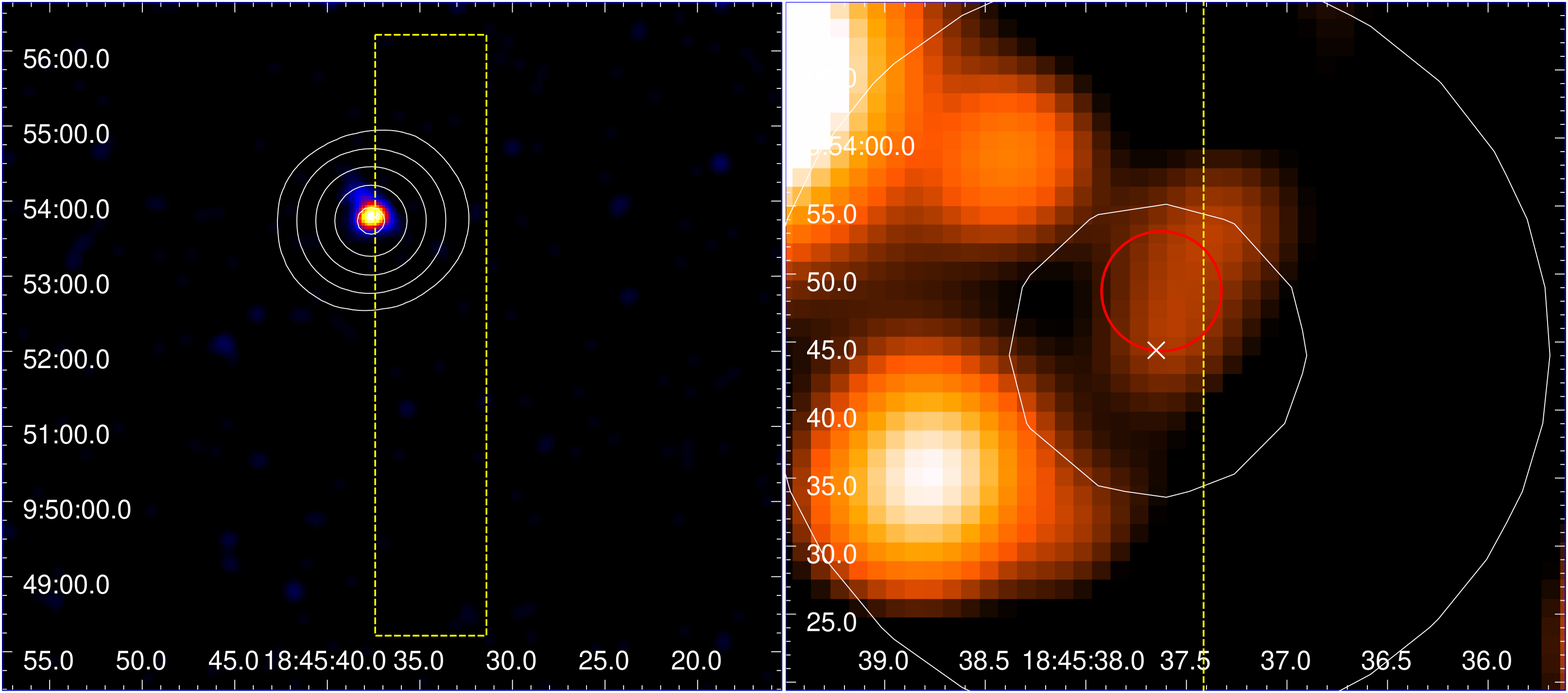}
\end{center}
\caption{Same as Fig.~\ref{fig:399} but for {\bf 3C~390}. White
  contours correspond in this case to 0.01, 0.1, 0.5, 1.5, and
  3~Jy~beam$^{-1}$.}
\label{fig:411}
\end{figure}

\subsubsection{3C 390}
\label{sub:411}

The X-ray source XRT~J184537.6$+$095349 ($S/N$=8.6) matches the radio
source NVSS~J184537$+$095344 in the field of view of 3C~390 at an
angular separation of 4.4~arcsec equal to the XRT error radius.
As shown in the left panel of Fig.~\ref{fig:411} the NVSS source is
not well centered with respect to the 3CR positional uncertainty
region.
The cross-match with the AllWISE Catalogue has provided as infrared
counterpart the source \wse~J184537.60$+$095345.0, at 0.9~arcsec from
the NVSS coordinates.
We note that there is another \wse source close to the one just
reported; both are not fully resolved in the image shown in the right
panel of Fig.~\ref{fig:411}.
However, the angular separation of the latter from the NVSS source is
higher (6.7~arcsec) and also larger than the established matching
radius (3.3~arcsec).
Moreover, we emphasise that \wse~J184537.60$+$095345.0 has been
recently included in the all-sky catalogue of blazar candidates by
\cite{2014ApJS..215...14D} due to its peculiar infrared colours.
Also in this case, no candidate to the radio source 3C~390 has been
previously presented in the literature: a spectroscopic analysis of
\wse~J184537.60$+$095345.0 will finally clarify the nature of this
multi-frequency source.

\subsubsection{3C 409}
\label{sub:414}

The soft X-ray source XRT~J201427.5$+$233455 has been detected with
$S/N$=11.6 in the field of view of 3C~409 and matches the coordinates
of NVSS~J201427$+$233452 with an angular separation of 1.9~arcsec.
X-ray emission in this region of the sky was indeed detected by the
Imaging Proportional Counter (IPC) on board the {\it Einstein}
satellite and reported by \cite{1983ApJ...269..400F}.
These authors also reported of a 7.3~ks exposure with the High
Resolution Imager (HRI) in which the X-ray source was located at
R.A.(B1950) 20$^h$ 12$^m$ 18$^s$.42, Dec.(B1950) $+$23$^{\circ}$
25$^{'}$ 45$^{''}$ with an uncertainty of 5~arcsec.
Considering the angular separation (5.8~arcsec) of this source from
XRT~J201427.5$+$233455 the X-ray emission detected from {\it Einstein}
and \sw is probably related to the same object.
The match with the AllWISE Source Catalogue has provided the infrared
conterpart \wse~J201427.59$+$233452.6 to the NVSS source, at only
0.3~arcsec from its coordinates; this infrared object is well detected
in all of the four \wse filters.
As in other three cases found in our analysis, this \wse infrared
object has been included in the all-sky catalogue of blazar-like
radio-loud sources recently produced by \cite{2014ApJS..215...14D}.
Apart from this, it is the first candidate ever indicated in the
literature as a possible counterpart to 3C~409.

\begin{figure}
\begin{center}
\includegraphics[height=3.9cm]{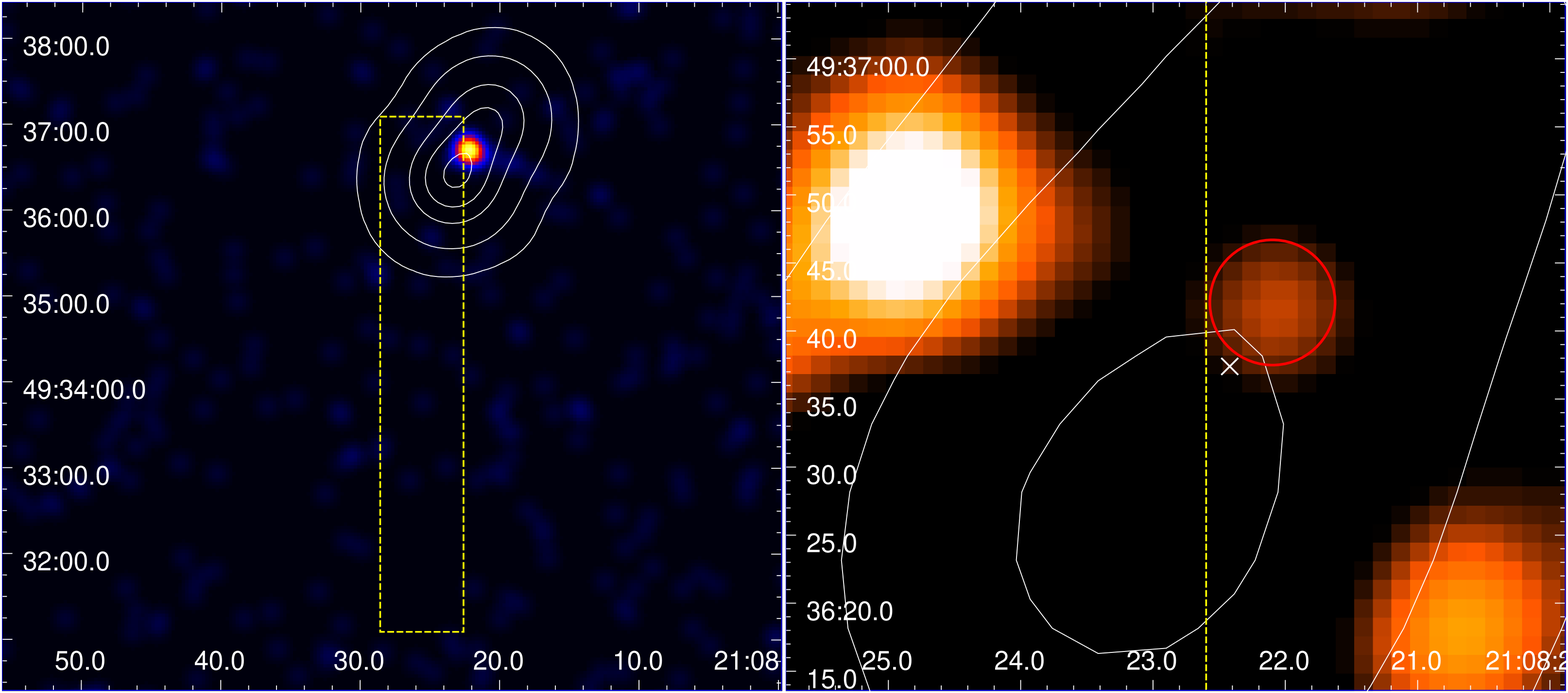}
\end{center}
\caption{Same as Fig.~\ref{fig:399} but for {\bf 3C~428}. White
  contours correspond in this case to 0.01, 0.1, 0.4, 0.7, and
  1~Jy~beam$^{-1}$.}
\label{fig:416}
\end{figure}

\subsubsection{3C 428}
\label{sub:416}

The X-ray source XRT~JJ210822.1$+$493642 matches the NVSS source
J210822$+$493637 in the field of view of 3C~428, at an angular
separation of 5.6~arcsec; the radio contours appear to be stretched in
one direction (see Fig~\ref{fig:416}).
As for 3C~86 and 3C~390 (see Fig.\ref{fig:399} and Fig.\ref{fig:411})
the coordinates of the NVSS source are not well centered with respect
to the 3CR positional uncertainty region: nonetheless, a substantial
overlap with its radio contours is evident.
From an infrared point of view a reliable infrared counterpart,
\wse~J210822.08$+$493641.6, has been found within the XRT error
circle: the angular separation from its center is only 0.5~arcsec,
much lower than the XRT error radius.
The image in the \wse $w1$ filter showing this infrared source is
given in the right panel of Fig.~\ref{fig:416}; the source is well
detected in all the remaining \wse filters.
As for 3C~91, 3C~390, and 3C~409, this infrared source has been
recently included in the all-sky catalogue of blazar candidates
\citep{2014ApJS..215...14D}.
Given the good match between the infrared and the X-ray source and
their low angular separation with respect to the NVSS coordinates, we
have finally accepted the match among these different sources
indicating \wse~J210822.08$+$493641.6 as the best target to be
investigated with a spectroscopic campaign.
We note that this position is at $\sim$~2~arcsec from one of the four
candidates (candidate B) suggested by \cite{1986JRASC..80..180H} in
their analysis of a CCD image obtained at the Canada-France-Hawaii
Telescope (CFHT); these authors claimed for this object a magnitude
$R$=21.8~mag.

\begin{figure}
\begin{center}
\includegraphics[height=3.9cm]{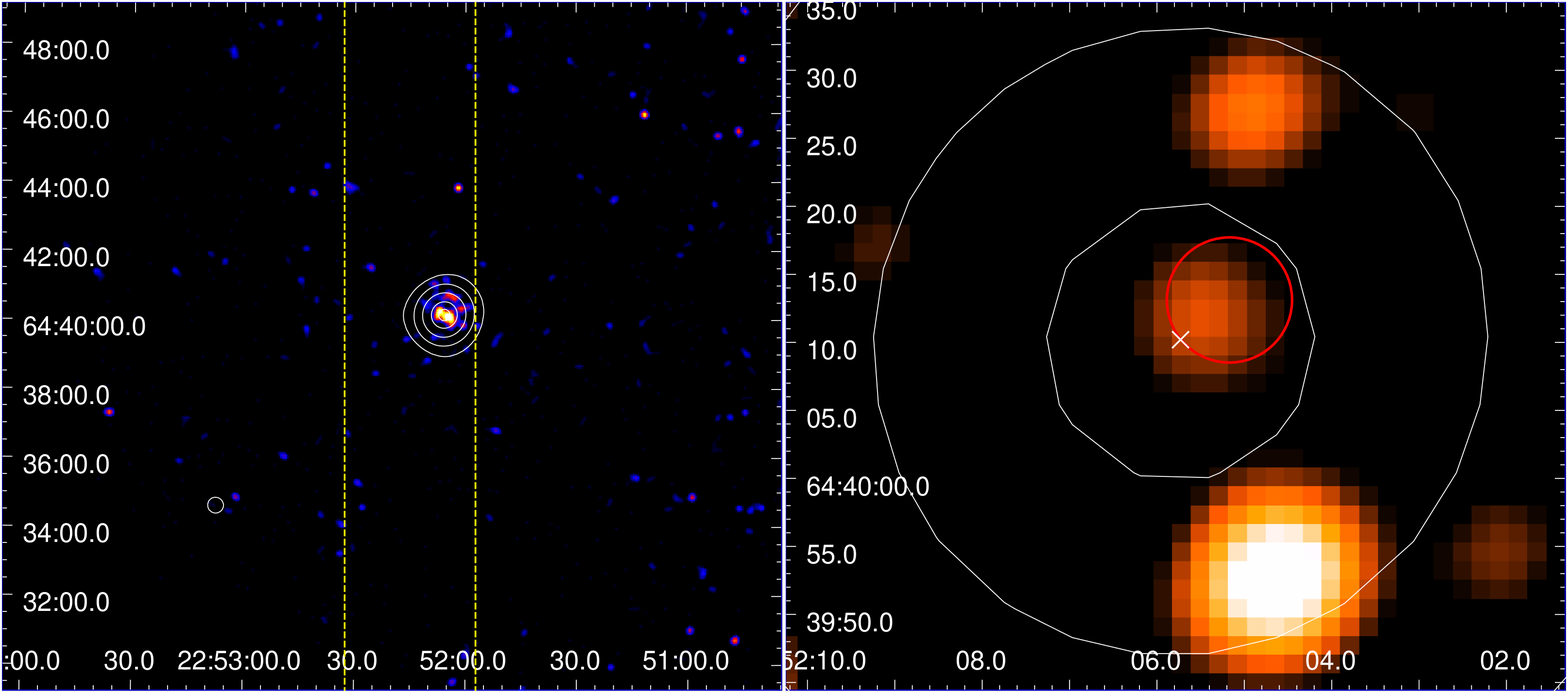}
\end{center}
\caption{Same as Fig.~\ref{fig:399} but for {\bf 3C~454.2}. White
  contours correspond in this case to 0.01, 0.1, 0.4, 1, and
  1.5~Jy~beam$^{-1}$.}
\label{fig:418}
\end{figure}

\subsubsection{3C 454.2}
\label{sub:418}

We have detected the soft X-ray source XRT~J225205.2$+$644013 at
4.6~arcsec from the coordinates of the NVSS source J225205$+$644010
within the positional uncertainty region of 3C~454.2, as shown in the
left panel of Fig.~\ref{fig:418}.
At an angular separation of 2.3~arcsec from this NVSS source,
consistent with the matching radius that we have established, the
infrared source \wse~J225205.50$+$644011.9 has been found in the
AllWISE Catalogue.
Excluding $w4$, it is well detected in all the remaining \wse filters,
as reported in Table~\ref{tab:lower}; the field of view in $w1$ is
shown in the right panel of Fig.~\ref{fig:418}.
No information about infrared or optical candidates was given before
in the literature about this radio source.

\subsection{3CR sources with only infrared counterparts}
\label{only_ir}

\begin{figure}
\begin{center}
\includegraphics[height=3.9cm]{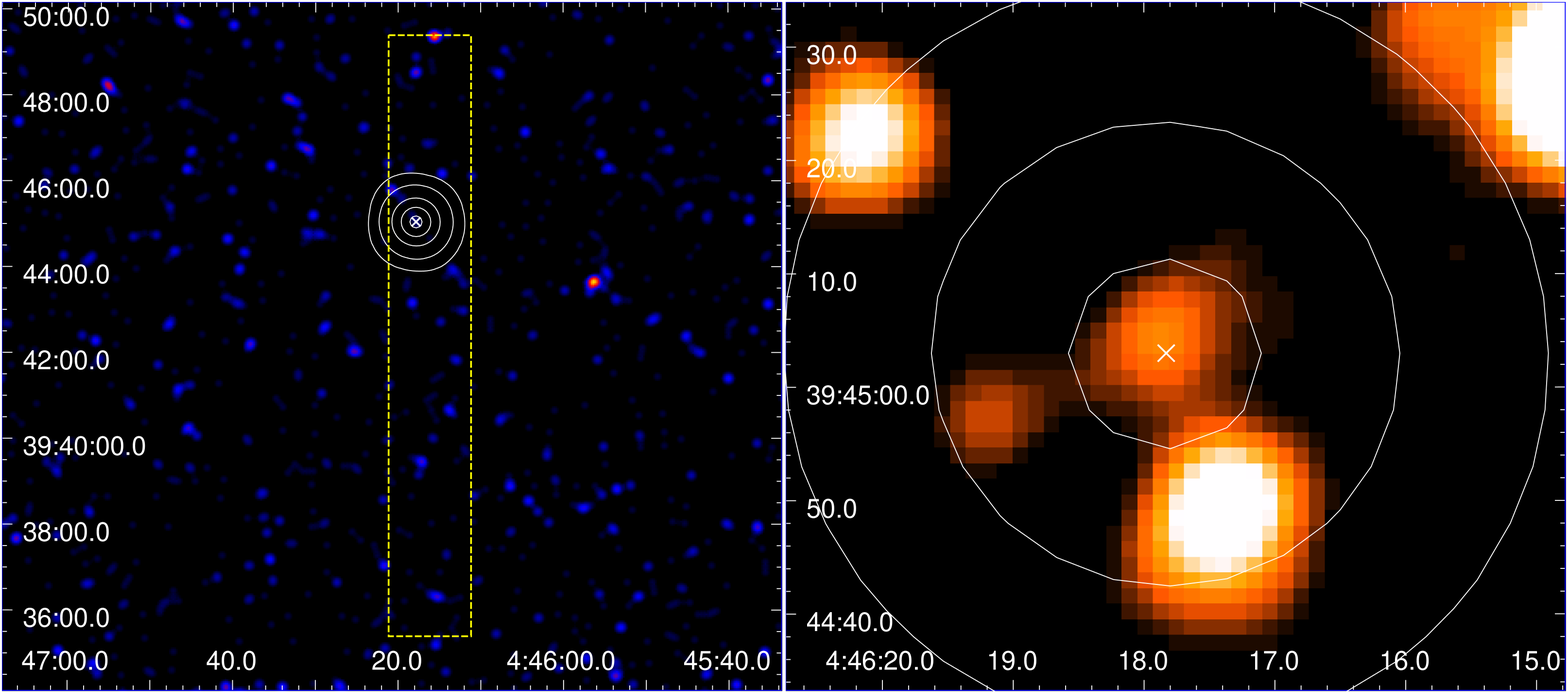}
\end{center}
\caption{The sky map in the direction of {\bf 3C~125} obtained by XRT
  in the 0.3--10 keV energy band (left panel) and by \wse in the $w1$
  filter (right panel). A yellow dashed line marks the 3CR positional
  uncertainty region. White continuous lines shape the radio contours
  obtained from the NVSS maps and corresponding to 0.01, 0.1, 0.5, 1,
  and 1.4~Jy~beam$^{-1}$; a white cross marks the position of the
  catalogued NVSS source.}
\label{fig:401}
\end{figure}

\subsubsection{3C 125}
\label{sub:401}

As shown in the XRT map in the 0.3--10 keV band (see
Fig.~\ref{fig:401}, left panel) no X-ray emission has been detected
for the source NVSS~J044617$+$394503, within the positional
uncertainty region of 3C~125.
From the cross-match with the AllWISE Catalogue, we have found the
source \wse~J044617.88$+$394504.5 with an angular separation of
1.6~arcsec from the coordinates of the NVSS source.
The infrared object is well detected in all four \wse filters; the
$w1$ image, given in the right panel of Fig.~\ref{fig:401}, shows good
match between the positions of the radio and infrared sources, without
evidence of confusion with other close objects.
No information has been found in the literature regarding
optical/infrared candidate counterparts for this radio source.

\begin{figure}
\begin{center}
\includegraphics[height=3.9cm]{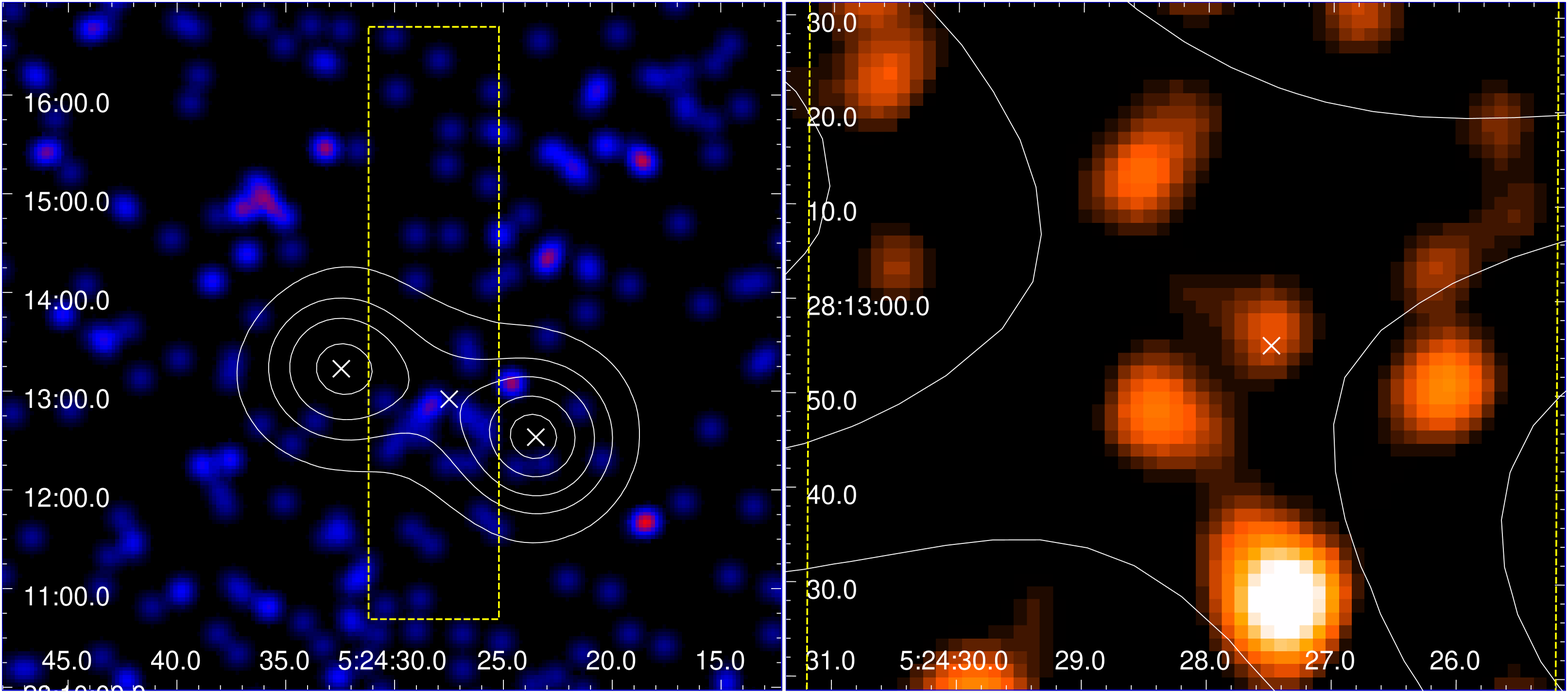}
\end{center}
\caption{Same as Fig.~\ref{fig:401} but for {\bf 3C~139.2}. White
  contours correspond in this case to 0.01, 0.08, 0.2, 0.4, and
  0.6~Jy~beam$^{-1}$.}
\label{fig:405}
\end{figure}

\subsubsection{3C 139.2}
\label{sub:405}

The radio contours of a source classified as FR~II
\citep{1984MNRAS.210..929L} overlap the positional uncertainty region
of 3C~139.2, and the NVSS object J050427$+$281255 is internal to it.
From the cross-match of the NVSS with the AllWISE Catalogue we have
obtained the infrared source \wse~J052427.51$+$281256.7, well detected
in all the four \wse filters, at an angular separation of 1.5~arcsec.
No information about infrared/optical identification has been found in
the literature for 3C~139.2.

\begin{figure}
\begin{center}
\includegraphics[height=3.9cm]{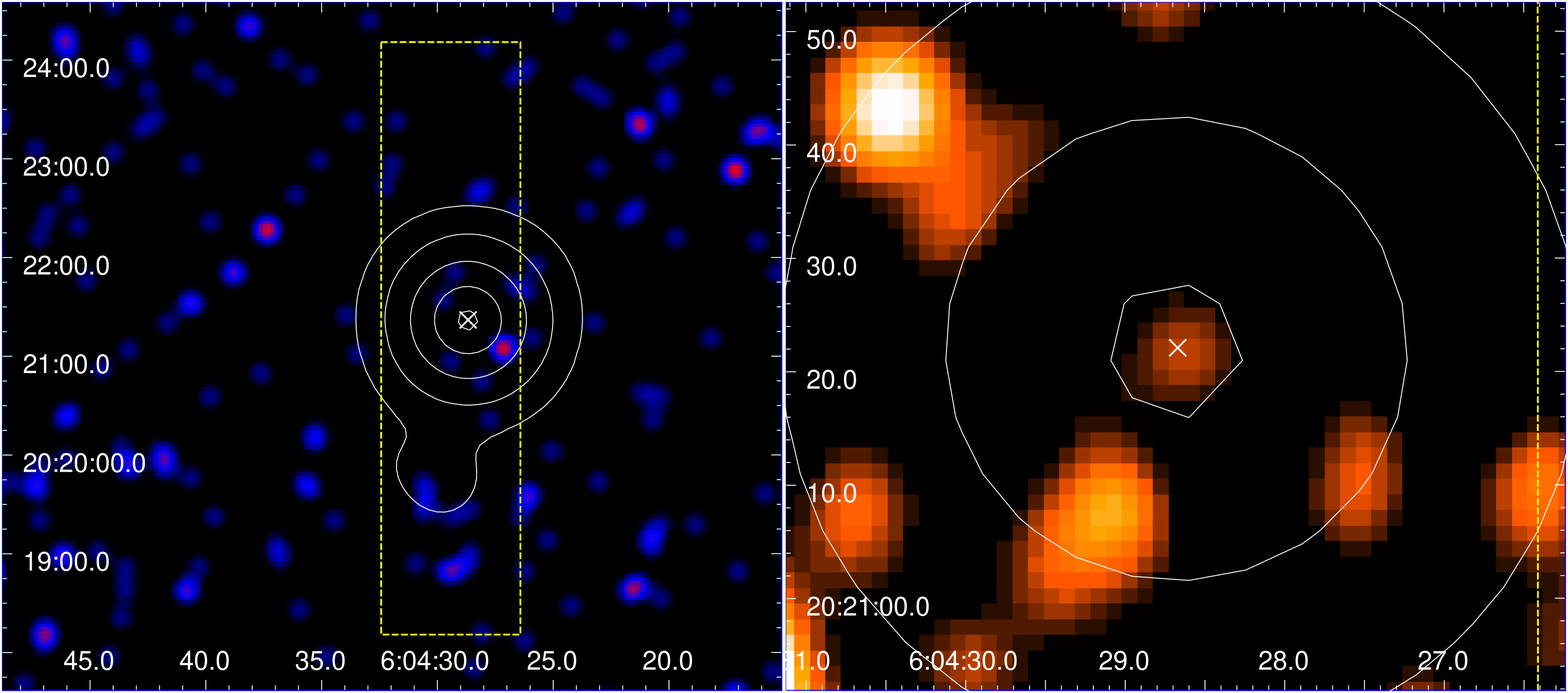}
\end{center}
\caption{Same as Fig.~\ref{fig:401} but for {\bf 3C~152}. White
  contours correspond in this case to 0.01, 0.1, 0.4, 0.9, and
  1.3~Jy~beam$^{-1}$.}
\label{fig:407}
\end{figure}

\subsubsection{3C 152}
\label{sub:407}

The left panel of Fig.~\ref{fig:407} shows the position of the source
NVSS~J060428$+$202122, at $\sim$15~arcsec from the center of the
positional uncertainty region of 3C~152, with no X-ray emission
detected by XRT.
From the cross-match with the AllWISE Catalogue we have found the
source \wse~J060428.62$+$202121.7, with an angular separation of only
0.8~arcsec between the two sources.
The infrared object is well detected in three of the \wse filters,
excluding $w4$.
The image in the $w1$ filter is given in the right panel of
Fig.~\ref{fig:407} and shows the good match between the positions of
the radio and the infrared objects.
In their analysis of the corresponding field of view with HST
\cite{1998AJ....115.1348M} reported a single candidate (R.A.(J2000)
06$^h$ 04$^m$ 28$^s$.63, Dec.(J2000) $+$20$^{\circ}$ 21$^{'}$
25$^{''}$.07).
The angular separation of this object from \wse~J060428.62$+$202121.7
is 3.4~arcsec.

\begin{figure}
\begin{center}
\includegraphics[height=3.9cm]{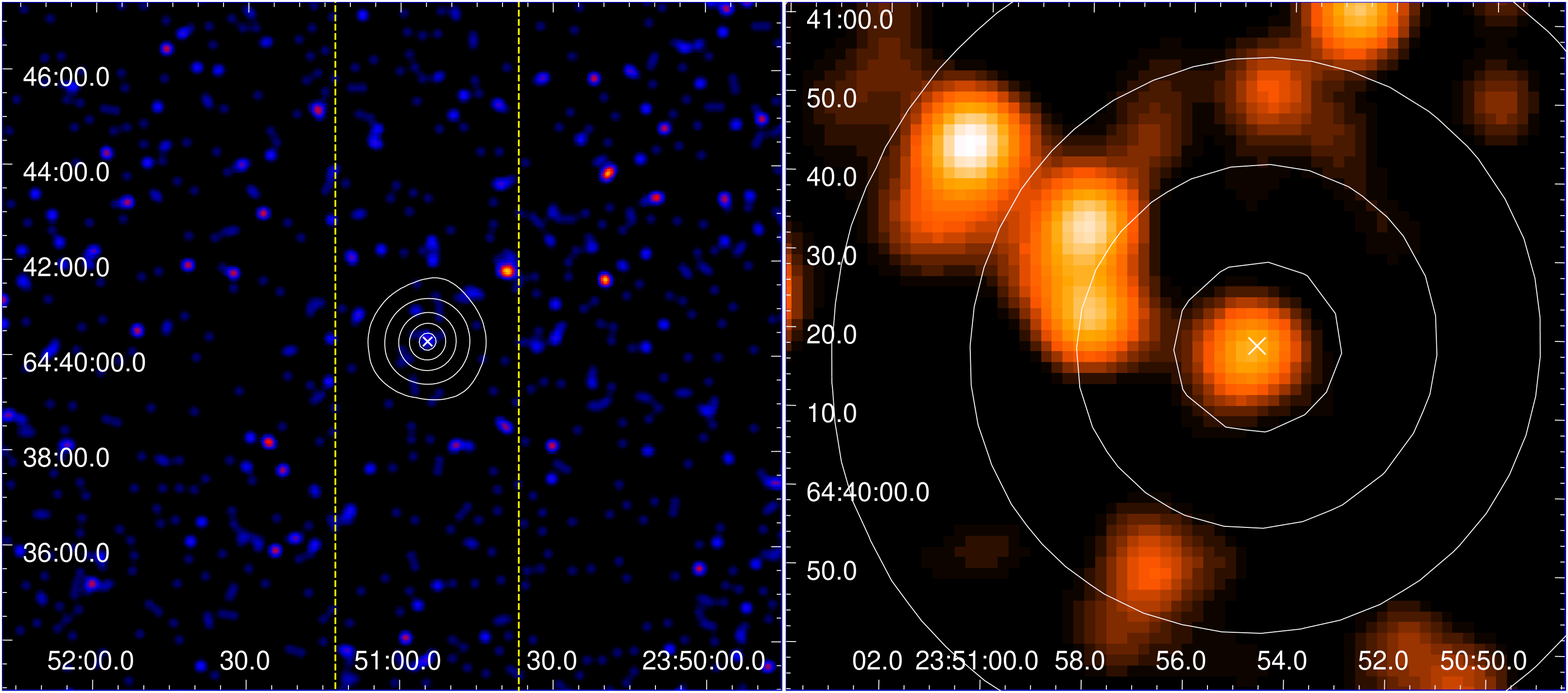}
\end{center}
\caption{Same as Fig.~\ref{fig:401} but for {\bf 3C~468.1}. White
  contours correspond in this case to 0.01, 0.2, 1.0, 2.2, and
  3.3~Jy~beam$^{-1}$.}
\label{fig:419}
\end{figure}

\subsubsection{3C 468.1}
\label{sub:419}

The left panel of Fig.~\ref{fig:419} shows the position of the 1.4~GHz
source NVSS~J235054$+$644018 with respect to the positional
uncertainty region of 3C~468.1.
The \wse candidate J235054.78$+$644018.1, that we have found from the
NVSS/AllWISE cross-match at an angular separation of 1.3~arcsec from
the NVSS source, is detected in all the four \wse filters.
\cite{1998AJ....115.1348M} presented the corresponding field of view
as observed by HST and reported a list of three tentative optical
identifications.
Their source \#1 (R.A.(J2000) 23$^h$ 50$^m$ 54$^s$.38, Dec.(J2000)
$+$64$^{\circ}$ 40$^{'}$ 18$^{''}$.06) is the closest (2.6~arcsec) to
the \wse source addressed by our analysis.
This value is consistent with the positional uncertainty (3~arcsec)
they reported for all their candidates.

\section{Summary and conclusions}
\label{sec:summary}

After conducting \sw observations of 21 bright NVSS sources
corresponding to 3CR sources classified as unassociated in the third
update of the 3CR catalogue, we have obtained new X-ray detections for
nine of them.
Moreover, cross-matching the NVSS with the recent AllWISE Catalogue,
we have found a \wse counterpart to all these nine X-ray sources, as
well as to four cases with no X-ray detection.
We have provided candidate counterparts emitting in the infrared band
for 3C~125, 3C~137, 3C~139.2, 3C~152, 3C~158, 3C~390, 3C~409, and
3C~454.2.
Furthermore, we have confirmed an unambiguous association for 3C~86,
3C~91, 3C~131, 3C~428, and 3C~468.1 where multiple candidates had been
suggested in previous analysis.
Four of these infrared sources are listed in the recent all-sky
catalogue of $\gamma$-ray blazar candidates
\citep{2014ApJS..215...14D}: the infrared colours of these objects are
similar to those of quasars
\citep{2011ApJ...740L..48M,2012ApJ...748...68D}, and only a
spectroscopic campaign will reveal the real nature of these as well as
of the remaining identified \wse counterparts.

It is worth mentioning that no optical/UV counterpart has been
detected in the UVOT filters at the position of the 21 NVSS sources:
this is in agreement with the notes reported in the 3CR catalogue
\citep{1985PASP...97..932S} in which the large fraction of these 3CR
unidentified radio sources were classified as {\it obscured} active
galaxies.
Therefore, our analysis suggests that a spectroscopic analysis in the
infrared range will be more helpful to identify their nature as well
as potentially obtain a redshift measurement.

\section*{Acknowledgements}

The authors are grateful to the anonymous referee for helpful comments
and suggestions. This work has been supported by ASI grant I/011/07/0.
This research has made use of archival data, software or online
services provided by the ASI Science Data Center; the High Energy
Astrophysics Science Archive Research Center (HEASARC) provided by
NASA's Goddard Space Flight Center; the SIMBAD database operated at
CDS, Strasbourg, France; the NASA/IPAC Extragalactic Database (NED)
operated by the Jet Propulsion Laboratory, California Institute of
Technology, under contract with the National Aeronautics and Space
Administration.  Part of this work is based on the NVSS (NRAO VLA Sky
Survey): The National Radio Astronomy Observatory is operated by
Associated Universities, Inc., under contract with the National
Science Foundation and on the VLA low-frequency Sky Survey (VLSS).
This publication makes use of data products from the Wide-field
Infrared Survey Explorer, which is a joint project of the University
of California, Los Angeles, and the Jet Propulsion
Laboratory/California Institute of Technology, and NEOWISE, which is a
project of the Jet Propulsion Laboratory/California Institute of
Technology. WISE and NEOWISE are funded by the National Aeronautics
and Space Administration.  SAOImage DS9 were used extensively in this
work for the preparation and manipulation of the images.

%%%%%%%%%%%%%%%%%%%%%%%%%%%%%%%%%%%%%%%%%%%%%%%%%%



\bibliographystyle{mnras}
\bibliography{3CR} % if your bibtex file is called example.bib

%\begin{thebibliography}{99}
%\end{thebibliography}

%%%%%%%%%%%%%%%%%%%%%%%%%%%%%%%%%%%%%%%%%%%%%%%%%%

%%%%%%%%%%%%%%%%% APPENDICES %%%%%%%%%%%%%%%%%%%%%

%\appendix



%%%%%%%%%%%%%%%%%%%%%%%%%%%%%%%%%%%%%%%%%%%%%%%%%%

% Don't change these lines
\bsp	% typesetting comment
\label{lastpage}
\end{document}